\documentstyle[prl,aps,psfig]{revtex}
\begin{document}
\draft
\twocolumn[\hsize\textwidth\columnwidth\hsize\csname
@twocolumnfalse\endcsname

\widetext
\title{Exact bounds on the ground-state energy of the infinite-$U$ Hubbard model}
\author{Federico Becca, Luca Capriotti, and Sandro Sorella}
\address{Istituto Nazionale per la Fisica della Materia and 
International School for Advanced Studies, \\
Via Beirut 4, I-34013 Trieste, Italy
}
\author{Alberto Parola}
\address{Istituto Nazionale per la Fisica della Materia and 
Dipartimento di Fisica, 
Universit\`a dell'Insubria, \\
Via Lucini 3, I-22100 Como, Italy
}
\date{\today}
\maketitle
\begin{abstract}
We give upper and lower bounds for the ground-state
energy of the infinite-$U$ Hubbard model. In two dimensions,
using these bounds we are able to rule out the possibility of
phase separation between the undoped-insulating state and an hole-rich state.
\end{abstract}
]
\narrowtext

The problem of phase separation (PS) in strongly correlated systems is one of
the most debated subject, especially after the seminal paper by Emery and co-workers
\cite{emery}, who also pointed out its relevance in the framework of high-temperature
superconductivity.
However, after several years of intensive numerical and analytical investigation,
there is no general consensus even for the qualitative features of the possible
instability in the Hubbard and $t-J$ models 
\cite{hellberg,calandra,cosentini,rommer,tandon,putikka}.

In this work we focus on the Hubbard model on a square lattice of 
$L$ sites in the infinite-$U$ limit,
and prove that there is no PS in the low doping limit: our result being
one of the few rigorous statements in this field.
The Hamiltonian is defined by
\begin{equation}
{\cal H} = -t \sum_{\langle i,j \rangle,\sigma} 
{\cal P}_G c^{\dag}_{i,\sigma} c_{j,\sigma} {\cal P}_G,
\end{equation}
where the $\langle i,j \rangle$ denotes nearest neighbor sites, and the 
Gutzwiller projector ${\cal P}_G$ 
enforces the constraint of no double occupancy on each site.

An upper bound on the ground-state (GS) energy can
be easily obtained by use of a variational wavefunction.
A very simple choice is given by the fully polarized ferromagnetic (FM)
state of energy 
$E_{FM}=2t\sum_{{\bf k}} \sum_{i=1,d} \cos k_i$,
where $d$ is the dimensionality of the lattice and 
the sum over ${\bf k}$ is restricted to the lowest hole energy orbitals.

A lower bound to the GS energy is instead more involved and can be derived as follows.
The GS of the Hamiltonian ${\cal H}$ with $M$ holes can be generally written as
\begin{equation}
|\psi \rangle =\sum_{{\bf R}_1 \ldots {\bf R}_M} 
\Psi({\bf R}_1 \ldots {\bf R}_M) 
|{\bf R}_1 \ldots {\bf R}_M \rangle,
\end{equation}
where $|{\bf R}_1 \ldots {\bf R}_M \rangle$ are normalized states 
representing suitable superpositions of spin states with the constraint
of having the holes placed in the sites labeled by  
$({\bf R}_1 \ldots {\bf R}_M)$.
The sum is over all distinct hole configurations.
The matrix elements $K_{ij}$ of the hopping operator $\hat K$ 
on the reduced Hilbert space defined by the hole configurations are 
non vanishing only if
a single hole hops at nearest neighbor distance. Therefore, $K_{ij}$ has no 
diagonal elements. For instance,
the element corresponding to the hopping ${\bf R}_1 \to {\bf R^\prime}_1$
is $\langle {\bf R^\prime}_1 \ldots {\bf R}_M |\hat{K} 
|{\bf R}_1 \ldots {\bf R}_M \rangle$ and its modulus is
always smaller or equal than $t$.
By construction, the lowest eigenvalue $E_0$ of this matrix is the
GS energy of the infinite-$U$ Hubbard model.

Now we prove that 
a lower bound of the GS energy is obtained by considering
the matrix with all the non vanishing entries equal to $-t$.
In fact, for the hermitian matrix with only non diagonal
elements $K_{ij}$, GS energy $E_0 <0 $ and with components
$\psi_i=\Psi({\bf R}_1 \ldots {\bf R}_M)$ we have:
\begin{eqnarray}
E_0&=& -\vert \sum_{i,j} \psi^*_iK_{ij}\psi_j \vert \ge 
-\sum_{i,j} \vert K_{ij}\vert\, |\psi_i|\, |\psi_j| \nonumber \\
&\ge& \sum_{i,j} B_{ij} \,|\psi_i| \,|\psi_j|\ge E_B
\label{frobenius}
\end{eqnarray}
Here, the matrix $B$ describes a gas of $M$ hard-core bosons 
(HCB) with nearest neighbor hopping 
$B_{ij}=-t$ for the non zero entries, which satisfies
$\vert B_{ij} \vert \ge  \vert K_{ij}\vert $.
The last inequality in (\ref{frobenius}) follows from
the Perron-Frobenius theorem \cite{mattis} and $E_B$ is the lowest 
eigenvalue of $B$.
This proves that a lower bound on
the energy of the infinite-$U$ Hubbard model is given by the GS
energy of an HCB gas with density equal to the doping
of the infinite-$U$ Hubbard model:  $E_B \le E_0 \le E_{FM} $.
These bounds clearly hold in any dimension $d$.

It is worth noting that the lower bound is not at all trivial, indeed
{\it i}) the Hilbert space for a system of $M$ HCB is much smaller than
the Hilbert space of a system of spin-1/2 with $M$ holes,
{\it ii}) in general it is not true that, for a given Hamiltonian
${\cal H}$, the energy of the bosonic GS is lower than the energy of the
fermionic GS. This is actually the case for the spin-1/2 infinite-$U$
Hubbard model but it does not hold in general (it is true if all the
off-diagonal matrix elements of ${\cal H}$ for the bosonic model are negative).
For example let us consider the two-dimensional $t-J$ model
\begin{eqnarray}
{\cal H}_{t-J} = -t \sum_{\langle i,j \rangle,\sigma} 
{\cal P}_G c^{\dag}_{i,\sigma} c_{j,\sigma} {\cal P}_G \nonumber \\
+J \sum_{\langle i,j \rangle} \left ( S_i \cdot S_j - \frac{1}{4}
n_i n_j \right ),
\label{tj}
\end{eqnarray}
where $S_i=\frac{1}{2} \sum_{\sigma,\sigma^{\prime}}
c^{\dag}_{i,\sigma} \tau_{\sigma,\sigma^{\prime}} c_{i,\sigma^{\prime}}$,
being $\tau_{\sigma,\sigma^{\prime}}$ the Pauli matrices,
and $n_i=\sum_{\sigma} c^{\dag}_{i,\sigma} c_{i,\sigma}$.
If we consider $L=16$ and $2$ holes, the 
fermionic GS has an energy lower than the corresponding bosonic GS \cite{note}
for $J\ge 0.2t$.
Indeed for this Hamiltonian, even for bosons, the off-diagonal terms are not
all negative definite. In Table I we report the GS energies of the $t-J$
model for different value of $J$ both for fermions and bosons.
Remarkably Lanczos diagonalizations on finite
clusters show that, even at finite $J$, the hole-hole correlations in the
$t-J$ model are quite similar to the corresponding HCB results \cite{becca}.

\begin{figure}
\centerline{\psfig{bbllx=30pt,bblly=200pt,bburx=560pt,bbury=590pt,%
figure=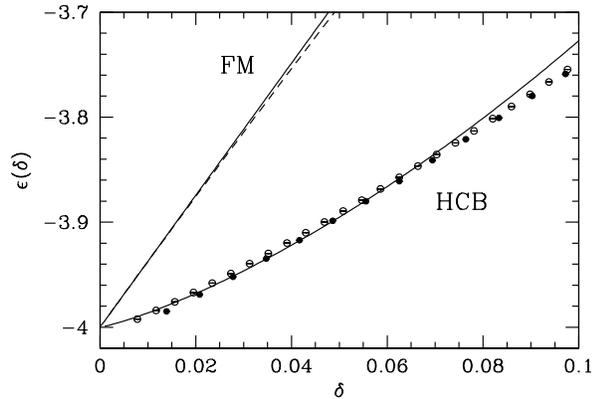,width=80mm,angle=0}}
\caption{\baselineskip .185in \label{fig1}
$\epsilon(\delta)$ for the FM state and for a gas of HCB.
Full and empty circles are Monte Carlo data for
$L=144$ and $L=256$, the dashed line is the analytical result for the
FM state and the continuous lines are the low-density asymptotic behaviors.
}
\end{figure}

Let us consider now the relevance of these bounds on the GS energy in the
two dimensional case of the infinite-$U$ Hubbard model.
The GS energy of an HCB gas can be computed 
numerically by quantum Monte Carlo on finite systems \cite{gfmc}. 
Moreover the T-matrix 
formalism in the low density limit allows to obtain the following analytic
expression \cite{lowdens}:
\begin{equation}
\frac{E_B(\delta)}{L} \simeq -4t \delta -2\pi t \frac{\delta^2}{\log \delta},
\end{equation}
with $\delta=M/L$. 

If PS occurs between a hole-free region (with $\delta=0$) and a hole-rich
phase (with $\delta=\delta_c$) then
for finite size systems the function
\begin{equation}
\epsilon(\delta)=\frac{E_0(\delta)-E_0(0)}{Mt}
\end{equation}
displays a minimum at $\delta=\delta_c$ 
\cite{emery} while, in the thermodynamic limit, $\epsilon(\delta)$ would be 
constant in the range $0\le\delta\le\delta_c$.

In Fig.~(\ref{fig1}) we show the previously discussed bounds on 
$\epsilon(\delta)$. 
Since $E_B(0)=E_{FM}(0)=E_0(0)=0$, $\epsilon(\delta)$
for the infinite-$U$ Hubbard model lies in between $\epsilon_B(\delta)$
and $\epsilon_{FM}(\delta)$.
For any $finite$ density the curvature of both curves is positive ruling
out the possibility to have PS starting from $\delta=0$.
Notice that a previous lower bound to the GS energy given by Trugman 
\cite{trugman} just corresponds to $\epsilon(\delta)=-4$, and then it 
would not allow to draw any conclusion on the occurrence of PS in this model.  

In conclusion we have obtained a rigorous result that rules out PS in the
infinite-$U$ Hubbard model starting from the zero doping limit.
Of course our findings do not exclude that PS could take place between two
finite densities or at finite $U>0$.

This work has been partially supported by INFM and MURST (COFIN99).

\begin{table}
\begin{tabular}{ccc}
$J/t$  & ${\cal E}_f/t$    & ${\cal E}_b/t$     \\
\hline \hline
0.1    & -7.1348    & -7.5695   \\
0.2    & -8.2118    & -8.1967   \\
0.3    & -9.4199    & -9.3755   \\
0.4    & -10.6839   & -10.6056  \\
0.5    & -11.9843   & -11.8897  \\
\end{tabular}
\caption{GS energies ${\cal E}_f$ and ${\cal E}_b$ for 2 holes on $L=16$ sites 
for the fermionic and bosonic $t-J$ model respectively.}
\end{table}
 

\end{document}